\documentclass[twocolumn,prl]{revtex4}
\usepackage{amsfonts}
\usepackage{graphicx}
\usepackage{amsmath}
\usepackage{amssymb}

\let\address=\affiliation

\begin{document}

\title{
Generation and Manipulation of Spin Current in Graphene Nanodisks:\\
Robustness against Randomness and Lattice Defects
}

\author{Motohiko Ezawa}

\address{Department of Applied Physics, University of Tokyo, Hongo 7-3-1, Tokyo 113-8656, Japan}

\begin{abstract}
Trigonal zigzag graphene nanodisk exhibits magnetism whose spin is proportional to the edge length of the nanodisk.
Its spin can be designed from 1/2 to a huge value. 
The spins form a quasiferromagnet, which has intermediate properties between a single spin and a ferromagnet.
In other words, the ferromagnet order has a relatively long life time,
 and yet the nanodisk spin can be rotated by external field or current.
We consider a nanodisk connected with two leads.
This system acts as a spin filter just as in a metal-ferromagnet-metal junction.
In this way we can generate a spin current.
Furthermore we can manipulate spin current by spin valve, spin switch and other spintronic devices
made of graphene nanodisks. We also show that nanodisk spins are robust against 
the effect of randomness in site energy, transfer energy and lattice defects. 
\end{abstract}

\maketitle


\textit{Introduction}: Graphene nanoribbons\cite{Fujita} have attracted
great interests in recent years due to their remarkable electronic and
magnetic properties and potential applications to nanoelectronic and
spintronic devices. Zigzag graphene nanoribbons show magnetism due to the
edge states though carbon itself is a nonmagnetic atom. Recently found
another structure, trigonal zigzag graphene nanodisk\cite%
{EzawaDisk,Fernandez,Wang}, also shows edge ferromagnetism.

Trigonal zigzag graphene nanodisks have degenerate zero-energy states in the
non-interacting regime\cite{EzawaDisk}. The inclusion of Coulomb
interactions provides the ground state with a finite spin proportional to
the edge length. The nanodisk-spin system undergoes a quasiphase transition
between the quasiferromagnet and the quasiparramagnet, which is a
finite-size version of a phase transition. By connecting a nanodisk with
leads, Coulomb blockades\cite{EzawaCoulomb} and Kondo effects\cite%
{EzawaKondo} have been predicted. In this paper we propose some application
of nanodisk to spintronics, such as a spin filter, a spin valve and a spin
switch. We also study if nanodisk spins are robust against randomness in
site and transfer energy and lattice defects.

\begin{figure}[t]
\begin{center}
\centerline{\includegraphics[width=0.4\textwidth]{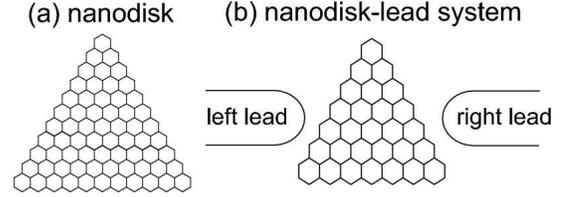}}
\end{center}
\caption{(a) Geometric configuration of a trigonal zigzag nanodisk. We
define its size $N$ by $N=N_{\text{ben}}-1$ with $N_{\text{ben}}$ the number
of benzenes on one side of the trigon. Here, $N=10$. \ (b)\ \ The
nanodisk-lead system. A nanodisk is connected to the right and left leads by
tunneling coupling.}
\label{FigNLeadY}
\end{figure}

\textit{Energy spectrum:} We calculate the energy spectrum of the nanodisk
based on the nearest-neighbor tight-binding model, which has proved to
describe accurately the electronic structure of graphene, carbon nanotubes,
graphene nanoribbons and other sp$_{2}$ carbon materials. The Hamiltonian is
defined by%
\begin{equation}
H_{0}=\sum_{i}\varepsilon _{i}c_{i}^{\dagger }c_{i}+\sum_{\left\langle
i,j\right\rangle }t_{ij}c_{i}^{\dagger }c_{j},  \label{HamilTB}
\end{equation}%
where $\varepsilon _{i}$ is the site energy, $t_{ij}$ is the transfer
energy, and $c_{i}^{\dagger }$ is the creation operator of the $\pi $
electron at the site $i$. The summation is taken over all nearest
neighboring sites $\left\langle i,j\right\rangle $. Owing to their
homogeneous geometrical configuration, we may take constant values for these
energies, $\varepsilon _{i}=\varepsilon _{\text{F}}$ and $t_{ij}=t\approx
2.70$eV. There exists one electron per one carbon, and the band-filling
factor is 1/2. Then, the diagonal term yields just a constant, $\varepsilon
_{\text{F}}N_{\text{C}}$, and can be neglected in the Hamiltonian, where $N_{%
\text{C}}$ is the number of carbon atoms. 
\begin{figure}[t]
\centerline{\includegraphics[width=0.5\textwidth]{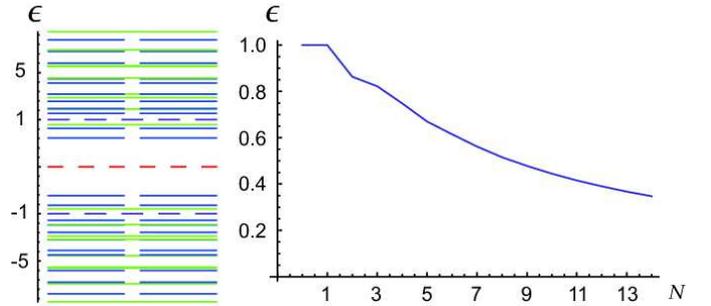}}
\caption{(a) Energy spectrum in the nanodisk with $N=6$. The vertical axis
is the energy $\protect\varepsilon $ in units of $t=2.7$eV. Segments in each
energy level indicate the degeneracy of the level. (b) The excitation gap $%
\protect\varepsilon $ as a function of the size $N$. It is approximately
proportional to $N^{-1}$.}
\label{gap}
\end{figure}

We show the excitation energy gap for the size-$N$ nanodisk in Fig.\ref{gap}%
. It has $N$-fold degenerate zero-energy states\cite{EzawaDisk}, where the
gap energy is as large as a few eV. Hence it is a good approximation to
investigate the electron-electron interaction physics only in the
zero-energy sector, by projecting the system to the subspace made of those
zero-energy states. The zero-energy sector consists of $N$ orthonormal
states $|f_{\alpha }\rangle $, $\alpha =1,2,\cdots ,N$, together with the SU(%
$N$) symmetry.

\textit{Quasiferromagnet:} We then introduce Coulomb interactions into the
zero-energy sector. The dominant contribution is well described by the
Hubbard model,

\begin{equation*}
H_{U}=U\sum_{i}c_{i\uparrow }^{\dagger }c_{i\uparrow }c_{i\downarrow
}^{\dagger }c_{i\downarrow },
\end{equation*}%
where it is estimated\cite{Pisani} that $U\approx t$. Within the model the
Coulomb energy $U_{\alpha \beta }$ and the exchange energy $J_{\alpha \beta
} $ between electrons in the states $|f_{\alpha }\rangle $ and $|f_{\beta
}\rangle $ are equal, $J_{\alpha \beta }=U_{\alpha \beta }$, and that all $%
J_{\alpha \beta }$ are of the same order of magnitude for any pair of $%
\alpha $ and $\beta $, implying that the SU($N$) symmetry is broken but not
so strongly\cite{EzawaCoulomb}. It is a good approximation to start with the
exact SU($N$) symmetry, by replacing $J_{\alpha \beta }$ with the average $J$%
, $J\approx 0.1U$. Then, the Hubbard model yields the infinite range
Heisenberg model,%
\begin{equation}
H_{\text{S}}=-J\sum_{\alpha \neq \beta }\mathbf{S}(\alpha )\cdot \mathbf{S}%
(\beta ),  \label{HamilIR}
\end{equation}%
where we have ignored irrelevant constant terms, and 
\begin{equation}
\mathbf{S}(\alpha )=\frac{1}{2}d_{\sigma }^{\dag }(\alpha )\mathbf{\tau }%
_{\sigma \sigma ^{\prime }}d_{\sigma ^{\prime }}(\alpha )
\end{equation}%
with $d_{\sigma }(\alpha )$ the annihilation operator of electron with spin $%
\sigma =\uparrow ,\downarrow $ in the state $|f_{\alpha }\rangle $: $\mathbf{%
\tau }$ is the Pauli matrix.

The total spin is given by $\mathbf{S}_{\text{tot}}=\sum_{\alpha }\mathbf{S}%
\left( \alpha \right) $. It takes the maximum value%
\begin{equation}
S_{g}=\sqrt{\frac{1}{2}N\left( \frac{1}{2}N+1\right) }
\end{equation}%
in the ground state, since all spins are spontaneously polarized into one
direction due to the exchange interaction. We refer the total spin $\mathbf{S%
}_{\text{tot}}$ in the ground state as the nanodisk spin.

We expect the nanodisk to be a ferromagnet provided $N\rightarrow \infty $.
It is actually not a rigid ferromagnet since the size $N$ is finite.
Nevertheless, the nanodisk spin system exhibits a strong ferromagnetic order
due to a large exchange interaction. The relaxation time is finite but quite
large even if the size $N$ is small. We have called such a system
quasiferromagnet\cite{EzawaDisk}.

A comment is in order. The spin of the ground state of the size-$N$ nanodisk
is $N/2$. This is consistent with Lieb's theorem\cite{Lieb} valid for the
Hubbard model. The theorem states that, in the case of repulsive
electron-electron interactions ($U>0$), a bipartite system at half-filling
has the ground state whose total spin is%
\begin{equation*}
S=\frac{1}{2}\left\vert N_{A}-N_{B}\right\vert ,
\end{equation*}%
where $N_{A}$ and $N_{B}$ are the numbers of sites in sublattices A and B,
respectively. Here, $N_{A}=(N+1)(N+6)/2$ and $N_{B}=(N+2)(N+3)/2$.

\textit{Effects of randomness:} In actual application, however, it is
important to discuss how stable the previous results are against lattice
defects and randomness in transfer energy. We study three types of
randomness: randomness in transfer energy, randomness in site energy and
lattice defects. The modified Hamiltonian is%
\begin{equation}
H_{0}=\sum_{i}\left( \varepsilon _{i}+\delta \varepsilon _{i}\right)
c_{i}^{\dagger }c_{i}+\sum_{\left\langle i,j\right\rangle }\left(
t_{ij}+\delta t_{ij}\right) c_{i}^{\dagger }c_{j},
\end{equation}%
where we take random values for $\delta \varepsilon _{i}$ and $\delta t_{ij}$%
. 

First of all the total spin of the ground state is determined by Lieb's
theorem although randomness is included. The total spin is given by the
difference of the A site and the B site 
\begin{equation}
S=\frac{1}{2}\left\vert N_{A}-N_{B}-\delta _{A}+\delta _{B}\right\vert
\end{equation}%
with the number of lattice defects at A (B) site $\delta _{A}$ ($\delta _{B}$%
), where $N_{A}$ and $N_{B}$ are number of A site and B site without lattice
defects. The total spin does not change by introducing randomness in
transfer and site energies. On the other hand, the number of the zero-energy
states changes by the number of lattice defects.

\begin{figure}[t]
\centerline{\includegraphics[width=0.3\textwidth]{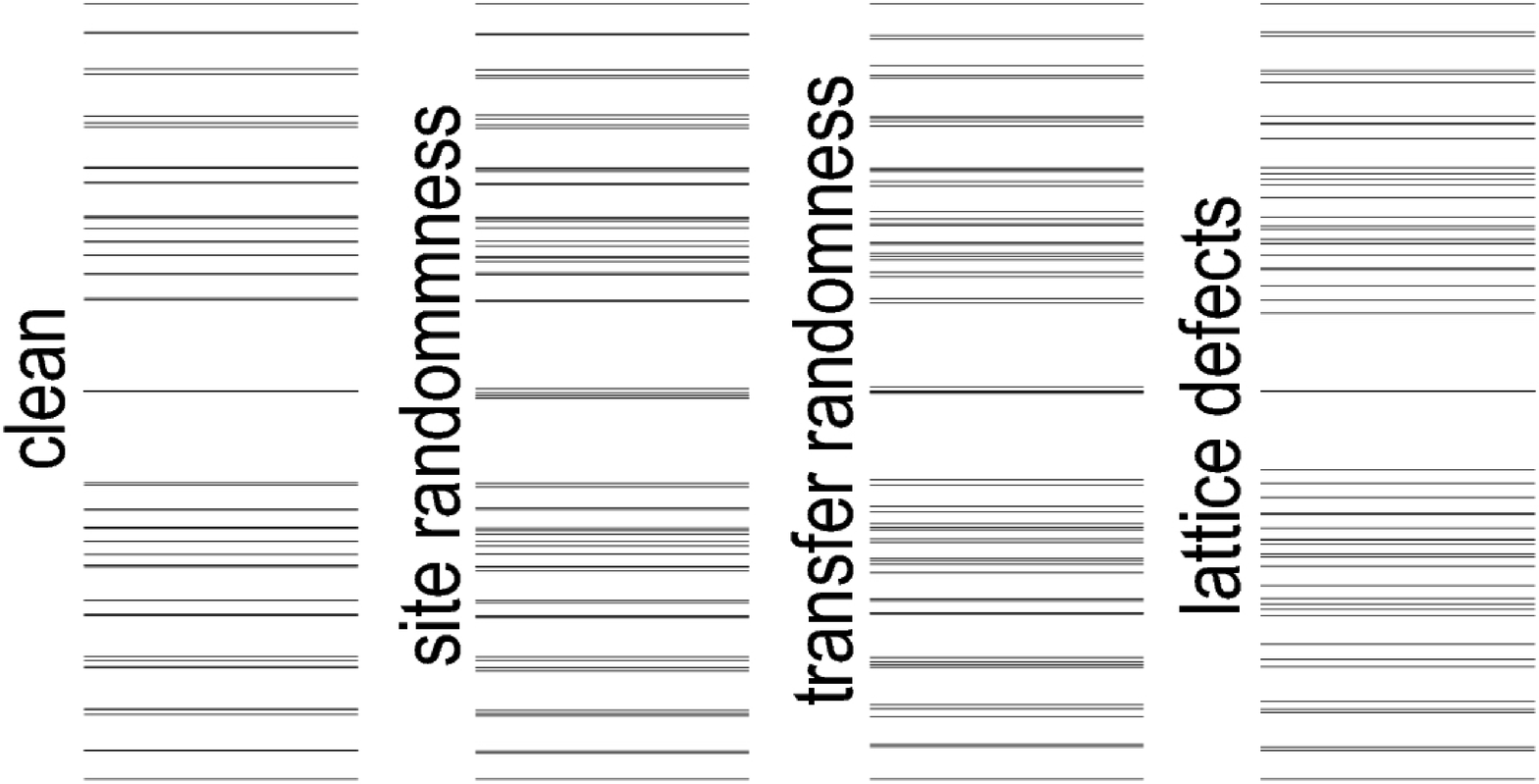}}
\caption{Energy level with randomness. (a) Clean nanodisks. (b) Site energy
randomness, $\protect\delta \protect\varepsilon _{i}/t\sim \pm 0.1$. (c)
Transfer energy randomness, $\protect\delta t_{ij}/t\sim \pm 0.1$. \ (d)
Lattice defects with thre site at a corner. See Fig.\protect\ref{FIgDefect}%
(a).}
\label{FigRandom}
\end{figure}

We show the energy spectrum with the randomness in Fig.\ref{FigRandom},
where we have taken rather large random values for $\delta \varepsilon _{i}$
and $\delta t_{ij}$: $|\delta \varepsilon _{i}|/t\leq \pm 0.1$ and $|\delta
t_{ij}|/t\leq \pm 0.1$. The zero energy remains as it is even when we
introduce lattice defects. However the zero-energy states split by the
randomness in transfer energy. The wave functions also change by the lattice
defect and the randomness in site and transfer energies. The changes are
proportional to the site (transfer) modification $\delta \varepsilon _{i}$ ($%
\delta t_{ij}$), but slight in site and transfer randomness. 

\begin{figure}[t]
\centerline{\includegraphics[width=0.4\textwidth]{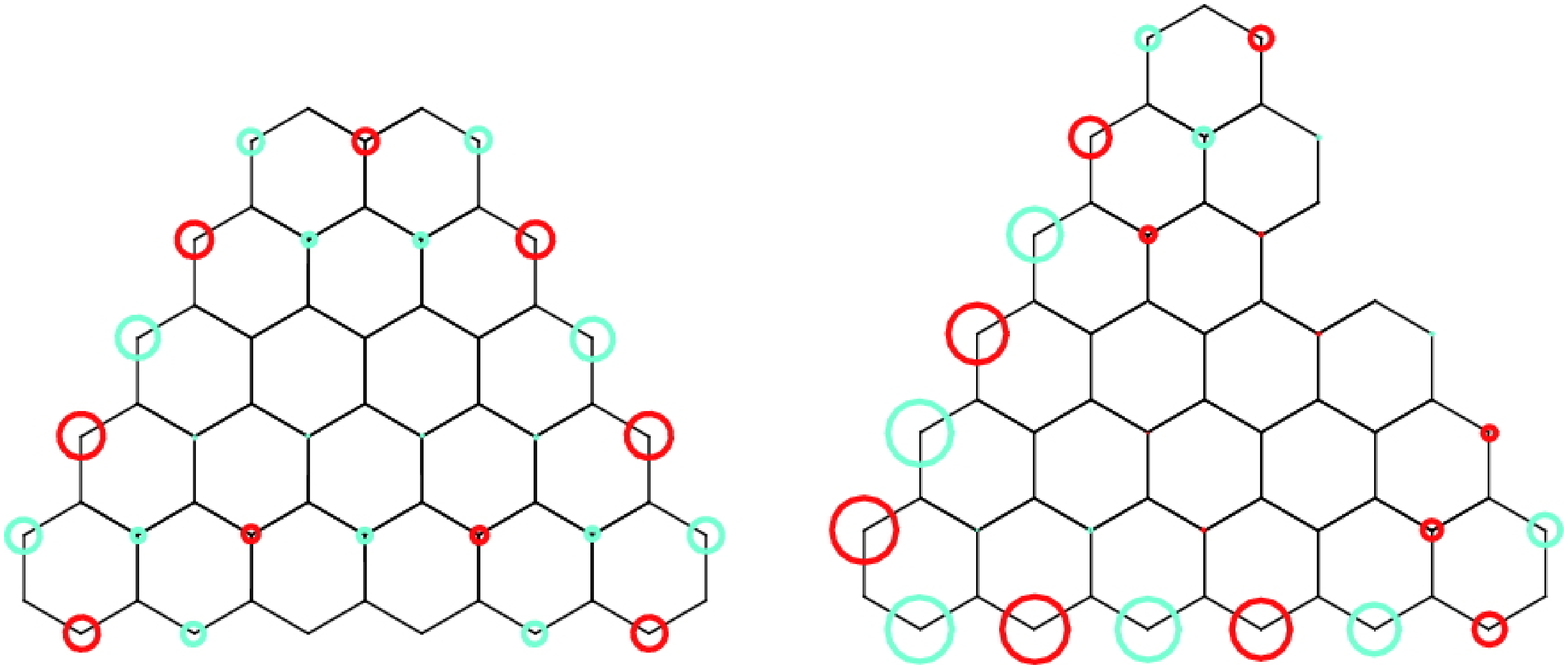}}
\caption{Wave function with lattice defects. (a) Three atoms are absent at
one corner. (b) One atom is absent at one edge. For both case the
probability density drastically reduces near the lattice defects. The number
of the zero-energy states reduces from $N$ to $N-1$ for both cases.}
\label{FIgDefect}
\end{figure}

On the other hand, the wave functions drastically change by introducing
lattice defects, as shown in Fig.\ref{FIgDefect}. We find the density
reduces drastically near the lattice defects. 

\textit{Application for spintronic devices:} The nanodisk-spin system is a
quasiferromagnet, which is an interpolating system between a single spin and
a ferromagnet. It is easy to control a single spin by a tiny current but it
does not hold the spin direction for a long time. On the other hand, a
ferromagnet is very stable, but it is hard to control the spin direction by
a tiny current. A nanodisk quasiferromagnet has an intermediate nature: It
can be controlled by a relatively tiny current\cite{EzawaSpin} and yet holds
the spin direction for quite a long time\cite{EzawaDisk}. Taking advantage
of these properties we propose some applications of graphene nanodisk-lead
systems to spintronic devices.

\begin{figure}[t]
\centerline{\includegraphics[width=0.3\textwidth]{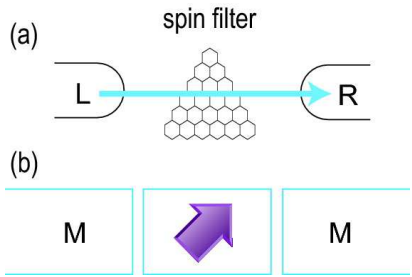}}
\caption{(a) An electron tunnels from the left lead to the nanodisk and then
to the right lead. The system is a reminiscence of a metal-ferromagnet-metal
junction. (b) Only electrons with the same spin direction as the nanodisk
spin can pass through the nanodisk freely. As a result, when we apply a
spin-unpolarized current to the nanodisk, the outgoing current is spin
polarized to the direction of the nanodisk spin. Consequently, this system
acts as a spin filter.}
\label{FigMFM}
\end{figure}

\textit{Spin filter:} We consider a lead-nanodisk-lead system [Fig.\ref%
{FigNLeadY}(b)], where an electron makes a tunnelling from the left lead to
the nanodisk and then to the right lead. This system is a reminiscence of a
metal-ferromagnet-metal junction (Fig.\ref{FigMFM}). If electrons in the
lead has the same spin direction as the nanodisk spin, they can pass through
the nanodisk freely. However, those with the opposite direction feel a large
Coulomb barrier and are blocked (Pauli blockade)\cite{EzawaSpin}. As a
result, when we apply a spin-unpolarized current to the nanodisk, the
outgoing current is spin polarized to the direction of the nanodisk spin.
Consequently, this system acts as a spin filter. 

\textit{Spin valve:} A nanodisk can be used as a spin valve, inducing the
giant magnetoresistance effect\cite{Fert,Grunberg}. We set up a system
composed of two nanodisks sequentially connected with leads (Fig.\ref%
{FigSpinValve}). We apply external magnetic field, and control the spin
direction of the first nanodisk to be $\left\vert \theta \right\rangle =\cos 
\frac{\theta }{2}\left\vert \uparrow \right\rangle +\sin \frac{\theta }{2}%
\left\vert \downarrow \right\rangle $, and that of the second nanodisk to be 
$\left\vert 0\right\rangle =\left\vert \uparrow \right\rangle $. We inject
an unpolarized-spin current to the first nanodisk. The spin of the lead
between the two nanodisks is polarized into the direction of $\left\vert
\theta \right\rangle $. Subsequently the current is filtered to the up-spin
one by the second nanodisk. The outgoing current from the second nanodisk is 
$I_{\uparrow }^{\text{out}}=I\cos \frac{\theta }{2}$. We can control the
magnitude of the up-polarized current from $0$ to $I$ by rotating the
external magnetic field. The system act as a spin valve. 
\begin{figure}[t]
\centerline{\includegraphics[width=0.38\textwidth]{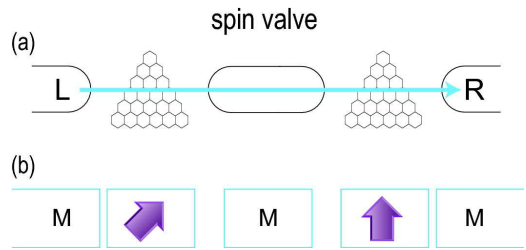}}
\caption{The spin valve is made of two nanodisks with the same size, which
are connected to leads. Applying an external magnetic field, we control the
spin direction of the first nanodisk to be $\left\vert \protect\theta %
\right\rangle $ and that of the second nanodisk to be $\left\vert \uparrow
\right\rangle $. The incoming current is unpolarized, but the outgoing
current is polarized and its magnitude can be controlled continuously. This
acts as a spin valve.}
\label{FigSpinValve}
\end{figure}

\textit{Spin switch:} We consider a chain of nanodisks and leads connected
sequentially (Fig.\ref{FigMFM}). Without external magnetic field, nanodisk
spins are oriented randomly due to thermal fluctuations, and a current
cannot go through the chain. However, when and only when a uniform magnetic
field is applied to all nanodisks, the direction of all nanodisk spins
become identical and a current can go through. Thus the system acts as a
spin switch, showing a giant magnetoresistance effect. The advantage of this
system is that a detailed control of magnetic field is not necessary in each
nanodisk. 
\begin{figure}[t]
\centerline{\includegraphics[width=0.5\textwidth]{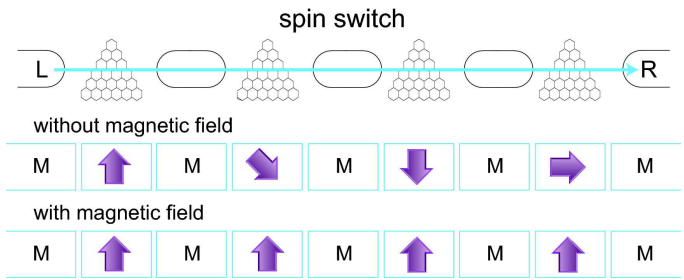}}
\caption{A chain of nonodisks and leads acts as a spin switch. Without
external magnetic field, nanodisk spins are oriented randomly due to thermal
fluctuations, and a current cannot go through the chain. However, as soon as
a uniform magnetic field is applied to all nanodisks, the direction of all
nanodisk spins become identical and a current can go through. }
\label{FigGMR}
\end{figure}

I am very much grateful to N. Nagaosa for many fruitful discussions on the
subject. This work was supported in part by Grants-in-Aid for Scientific
Research from the Ministry of Education, Science, Sports and Culture No.
20940011.

\end{document}